\documentclass[preprint]{aastex631}

\newcommand{\alfven}{Alfv\'{e}n}

\begin{document}

\title{Quadrupole White-light Sources in an X1.2 Flare Observed by ASO-S/LST/WST and SDO/HMI}

\author[0000-0002-8401-9301]{Zhichen Jing}
\author[0000-0002-8258-4892]{Ying Li}
\affiliation{Key Laboratory of Dark Matter and Space Astronomy, Purple Mountain Observatory, CAS, Nanjing 210023, People's Republic of China}
\affiliation{School of Astronomy and Space Science, University of Science and Technology of China, Hefei 230026, People's Republic of China}

\author[0000-0003-0057-6766]{Dechao Song}
\author[0000-0001-7540-9335]{Qiao Li}
\affiliation{Key Laboratory of Dark Matter and Space Astronomy, Purple Mountain Observatory, CAS, Nanjing 210023, People's Republic of China}

\author{Zhengyuan Tian}
\affiliation{Key Laboratory of Dark Matter and Space Astronomy, Purple Mountain Observatory, CAS, Nanjing 210023, People's Republic of China}
\affiliation{School of Astronomy and Space Science, University of Science and Technology of China, Hefei 230026, People's Republic of China}

\correspondingauthor{Ying Li, Dechao Song}
\email{yingli@pmo.ac.cn, dcsong@pmo.ac.cn}

\begin{abstract}
We present observations of an X1.2 white-light flare on 2023 January 6, which exhibits a rare quadrupolar white-light source configuration. This event was observed by the White-light Solar Telescope (WST; 3600~\AA) aboard the Advanced Space-based Solar Observatory and the Helioseismic and Magnetic Imager (HMI; 6173~\AA) aboard the Solar Dynamics Observatory. Four flare-related footpoints (labeled as FP1--FP4) were nearly simultaneously identified in both WST 3600~\AA\ and HMI 6173~\AA\ continua, associated with a quadrupolar magnetic configuration and a failed filament eruption. The inner sources of FP1 and FP2 showed a similar enhancement of $\sim$65\%/10\% in the WST/HMI continuum, while the outer sources of FP3 and FP4 exhibited weaker responses. The inner footpoints had earlier responses in UV and EUV bands and were spatially coincident with the hard X-ray (HXR) footpoint sources. The two southern footpoints (FP2 and FP4) showed stronger HXR and white-light emissions than their northern counterparts (FP1 and FP3), with FP4 uniquely exhibiting a distinct HXR emission above 60~keV, in contrast to the absence of such an emission at FP3. Notably, faint WST 3600~\AA\ enhancements at FP4 were observed during the gradual phase, temporally consistent with the fallback of filament material. This X1.2 flare presents a novel quadrupolar white-light structure, enriching our understanding of the generation and evolution of white-light flares.
\end{abstract}


\section{Introduction} 
\label{sec:intro}

Solar flares are among the most energetic phenomena in the solar atmosphere with sudden brightenings in the whole electromagnetic spectrum \citep{2011SSRv..159...19F}. The CSHKP model \citep{1964NASSP..50..451C, 1966Natur.211..695S, 1974SoPh...34..323H, 1976SoPh...50...85K} is one of the most famous flare models and the foundation of the standard flare model, which is often associated with a filament eruption \citep{1995ApJ...451L..83S}. Magnetic reconnection is the key process in this standard model and can release substantial energy and energetic particles. The upward-propagating non-thermal particles may contribute to plasma heating above the reconnection site, potentially facilitating the upward motion of the filament, which could form the core of a coronal mass ejection (CME). Meanwhile, the downward-propagating particles can heat the loop top and form the hard X-ray (HXR) loop-top source. At the chromosphere, the accelerated electrons can further produce nonthermal bremsstrahlung radiation, leading to the HXR flux enhancement and the HXR footpoint sources \citep{2007ApJ...654..665T}. A quadrupolar field can consist of a sheared core field and an overlying field, with three main triggering mechanisms: tether-cutting reconnection, breakout reconnection, and ideal magnetohydrodynamics (MHD) instability or loss of equilibrium \citep{2006GMS...165...43M}. The key difference between the former two scenarios is the sequence of reconnections: the tether-cutting starts with the inner reconnection in the sheared core field, while the breakout starts with the outer reconnection at the null point. As for the third scenario, there is no reconnection at the null point or in the core field before the eruption, and the reconnection may be triggered subsequently.  

White-light flares (WLFs) are a type of solar flare with an enhancement in the visible continuum or integrated light \citep{1970SoPh...13..471S, 1989SoPh..121..261N}. It is found that the white-light (WL) emission usually has good temporal and spatial correlations with the HXR emission in many WLFs \citep[e.g.,][]{2006SoPh..234...79H, 2011ApJ...739...96K, 2016ApJ...816....6K}. WL sources generally coincide with the HXR footpoint sources, with their peak times typically occurring around the same time during the rise phase of the flare \citep{2007ApJ...656.1187F, 2025ApJ...983L..41S}. In recent years, various possible heating mechanisms have been proposed: electron beams \citep{1972SoPh...24..414H,2007ApJ...656.1187F,2020ApJ...891...88W}, proton beams \citep{1978SoPh...58..363M}, irradiation from X-ray \citep{1978SoPh...60..341M} or extreme ultraviolet (EUV) \citep{1988SoPh..115..277P}, chromospheric backwarming \citep{2003A&A...403.1151D}, chromospheric condensation \citep{1992ApJ...397..694G, 2000A&A...354..691G}, and \alfven\ waves \citep{1982SoPh...80...99E}. Multiple mechanisms could contribute to the continuum emission in a certain flare \citep[e.g.,][]{2017NatCo...8.2202H,2023ApJ...952L...6S}.

Despite the large number of solar flares, only a small fraction have been identified as WLFs. Only about 100 WLFs had been observed before 2010 from ground-based and space-based observations \citep{2010ApJ...711..185C}. Fortunately, the situation has significantly improved with the advent of several space missions. The Helioseismic and Magnetic Imager \citep[HMI;][]{2012SoPh..275..207S} aboard the Solar Dynamics Observatory \citep[SDO;][]{2012SoPh..275....3P} can provide full-disk images in a narrow wave band around Fe~\textsc{i} 6173 \AA\ which has been widely used as a proxy for the white-light emission in the Paschen continuum \citep[e.g.,][]{2014ApJ...780L..28M, 2016ApJ...816....6K, 2026ApJ..1001..195S}. The Advanced Space-based Solar Observatory \citep[ASO-S;][]{2023SoPh..298...68G} is the first comprehensive solar mission in China, and it has three payloads: the Full-disk vector MagnetoGraph \citep[FMG;][]{2019RAA....19..161S}, the Hard X-ray Imager \citep[HXI;][]{2019RAA....19..160Z,2024SoPh..299..153S}, and the Lyman-alpha Solar Telescope \citep[LST;][]{2019RAA....19..158L, 2019RAA....19..162F, 2024SoPh..299..118C}. LST has three scientific instruments \citep{2019RAA....19..159C}: a Solar Disk Imager (SDI), a Solar Corona Imager (SCI), and a White-light Solar Telescope (WST). WST can provide full-disk images in the 3600$\pm$20 \AA\ wave band in the Balmer continuum since October 2022. With the instruments of HMI and WST, more WLFs have been observed: over 100 WLFs have been reported by HMI to date \citep[e.g.,][]{2018ApJ...867..159S, 2020ApJ...904...96C, 2026ApJ..1001..195S} and about 300 WLFs have been observed by WST till June 2024 \citep{2025ApJ...992...72J}.

The WL brightening can be observed both on the solar disk and off the solar limb, and both types of brightening are important for improving our understanding of morphological structure and physical mechanisms of WL emissions. The first observed flare, i.e., the Carrington event in 1859 \citep{1859MNRAS..20...13C, 1859MNRAS..20...15H}, was indeed a WLF on the solar disk. Unlike the intense brightness of this famous WLF, the WL enhancement in many flares is subtle and requires alternative identification methods, such as difference image analysis. This problem is more serious for the off-limb (i.e., above-the-limb) WL brightening, which may be affected by noise fluctuations, scattered light, and other observational limitations. Even so, some flares with off-limb WL brightenings have still been observed by ground-based telescopes \citep{1992PASJ...44...55H}. \cite{2014ApJ...780L..28M} reported the chromospheric and coronal WL sources in two X-class flares with HMI data, observed for the first time from a telescope in space. To date, nearly 100 flares with off-limb WL features have been reported. These events have been classified into several types based on their observational characteristics. For example, \citet{2023A&A...672A..32F} categorized the off-limb events into ejection, loop, and spike types according to their morphological appearance in WL images, while \citet{2023SoPh..298..148Z} further classified them into closed-loop eruptions, open-loop eruptions, fast ejections, and flare arcades by considering not only morphology but also their dynamical evolution and magnetic configurations. It should be noted that these classification schemes are not strictly unified and may overlap in some cases. HMI observes off-limb WL emissions in the Paschen continuum while WST provides observations in the Balmer continuum. \cite{2024ApJ...963L...3L} presented off-limb 3600 \AA\ emission in the Balmer continuum using WST data in an X1.5 flare whose WL emission is supposed to be caused by thermal plasma cooling or Thomson scattering.  

This paper focuses on an X1.2 flare (SOL2023-01-06T00:57) that occurred near the solar limb, featuring prominent WL brightenings associated with a rare quadrupolar configuration. The observational data are described in Section~\ref{sec:data}. The results including WL and multiwavelength analyses are presented in Section~\ref{sec:results}, and a summary and discussions are provided in Section~\ref{sec:summary}.

\section{Observational Data} 
\label{sec:data}

ASO-S/LST/WST provides WL observations in the 3600$\pm$20 \AA\ wave band, covering a field of view of 1.2 R$_\odot$. It has a pixel size of $\sim$0.5$^{\prime\prime}$ and a spatial resolution of about 4$^{\prime\prime}$ due to the point spread function. During this flare event, WST operated in the routine mode with a cadence of 2 min. ASO-S/HXI observes X-rays in the range of $\sim$10--300 keV, providing both spectra and imaging with a best spatial resolution of 3.1$^{\prime\prime}$ and a temporal resolution of up to 0.125~s \citep{2024SoPh..299..153S}. This work utilizes the G3--G10 sub-detector groups of HXI, providing a spatial resolution of 6.5$^{\prime\prime}$, and the HXI light curves are temporally rebinned to a cadence of 4~s.

SDO/HMI provides full-disk images of the pseudo-continuum at Fe~\textsc{i} 6173 \AA\ and also magnetograms with a cadence of 45~s and a spatial resolution of about 1$^{\prime\prime}$. Since 2013, HMI has been capable of observing an annular region extending $\sim$40$^{\prime\prime}$ beyond the solar limb \citep{2023A&A...672A..32F}. The HMI continuum intensity is reconstructed from filtergrams obtained at six wavelength positions across the \ion{Fe}{1}~6173~\AA\ line using a Michelson Doppler Imager \citep[MDI;][]{1995SoPh..162..129S}-like algorithm \citep{2016SoPh..291.1887C}. Previous studies have shown that this reconstruction performs well in quiet-Sun and moderate flare conditions, but the reliability may decrease in regions with strong magnetic fields or when the line profile deviates significantly from absorption \citep[e.g.,][]{2015SoPh..290..689C, 2018ApJ...860..144S}. For the present event, we checked all the available \ion{Fe}{1} profiles and found that they all remain in absorption and exhibit an overall intensity increase due to the enhanced continuum to some extent (see Appendix \ref{sec:appendix} and the associated figure) and the magnetic field strength in the WL sources are not greater than 900~G. Therefore, the reconstructed HMI continuum is supposed to be reliable for the flare under study. 

 SDO also carries the Atmospheric Imaging Assembly \citep[AIA;][]{2012SoPh..275...17L}, which provides full-disk images with a field of view of 1.5 R$_\odot$. AIA observes in seven EUV bandpasses sensitive at different temperatures: 94 \AA\ ($10^{6.8}$~K), 131 \AA\ ($10^{5.6}$, $10^{7.0}$~K), 171 \AA\ ($10^{5.8}$~K), 193 \AA\ ($10^{6.2}$, $10^{7.3}$~K), 211 \AA\ ($10^{6.3}$~K), 304 \AA\ ($10^{4.7}$~K), and 335 \AA\ ($10^{6.4}$~K). Additionally, it has two ultraviolet (UV) bandpasses at different characteristic temperatures: 1600 \AA\ ($10^{5.0}$~K) and 1700 \AA\ ($10^{3.7}$~K). The pixel size of AIA is 0.6$^{\prime\prime}$, with a spatial resolution of 1.5$^{\prime\prime}$. The cadences of EUV and UV observations are 12 and 24 s, respectively. In this work, we use both EUV and UV images from AIA. Besides, the Geostationary Operational Environmental Satellite (GOES; \citealt{2004SPIE.5570..155K}) provides the soft X-ray (SXR) 1--8 \AA\ data via its X-ray Sensor (XRS) instrument \citep{1996SPIE.2812..344H} for the flare under study.

\section{Results} 
\label{sec:results}

\subsection{Overview of the X1.2 White-light Flare} 
\label{subsec:overview}

\begin{figure*}[ht!]
\centering
\includegraphics[width=0.75\textwidth]{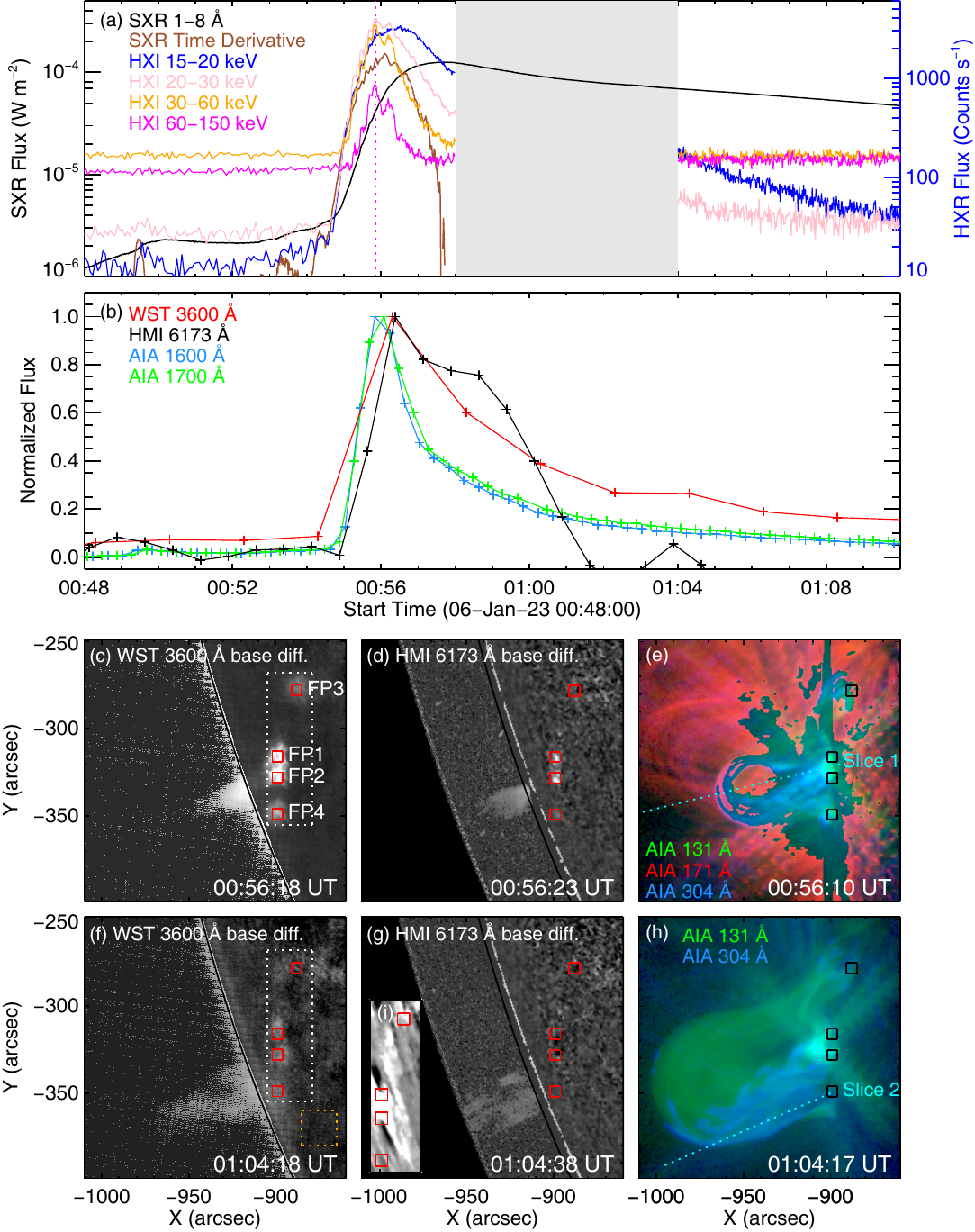}
\caption{(a) Light curves of GOES SXR 1--8 \AA, SXR time derivative, and HXR emissions at different energy bands. The magenta dotted line represents the peak time of the HXI 60--150 keV emission. The light gray shaded area marks the period when HXI emissions were affected by the radiation belt. (b) Normalized emission curves of WST 3600 \AA, HMI 6173 \AA, and AIA UV 1600 and 1700 \AA\ summed over the flare region indicated by the white dotted box in panels (c) and (f), with pre-flare background subtracted. (c)--(e) Multiwavelength images at the flare peak time, with the solar limb outlined by the black curve. Note that the off-limb parts in the WST and HMI base-difference images are logarithmically enhanced, and the flare core region in the 171 \AA\ image (panel (e)) is affected by saturation. Four red or black boxes in each panel outline the footpoint-like sources as detected in the WST and HMI images, labeled as FP1--FP4. (f)--(h) Same as the images in panels (c)--(e) but during the decay phase of the flare. The orange dotted box in panel (f) marks the quiet-Sun region used to estimate the measurement uncertainties of the WST and HMI emission enhancements. The cyan dotted lines in panels (e) and (h), referred to as Slice~1 and Slice~2, respectively, are used to plot the time-distance diagrams as shown in Figures \ref{fig:lc12} and \ref{fig:lc34}. (i) HMI magnetogram of the region indicated by the white dotted box in panels (c) and (f), obtained one hour later (02:04:38 UT).
\label{fig:1}}
\end{figure*}

The X1.2 flare on 2023 January 6 was located in NOAA active region 13182 (S20E81), near the southeastern solar limb. According to the GOES SXR 1--8 \AA\ observation, the flare began at 00:43~UT, peaked at 00:57~UT, and ended at 01:07~UT. There was a small bump in the SXR 1--8 \AA\ flux at $\sim$00:50~UT in the early rise phase of the flare but no notable HXR enhancement was detected (see Figure \ref{fig:1}(a)). After $\sim$00:55~UT, the SXR emission began to increase rapidly, indicative of a rapid release of magnetic energy. The HXI 15--20~keV flux reached its maximum at $\sim$00:57~UT and subsequently exhibited a gradual decay, resembling the trend of the SXR 1--8 \AA\ emission. At the higher energy bands of 20--30, 30--60, and 60--150~keV, the HXI fluxes peaked slightly earlier at 00:56~UT (see the magenta vertical dotted line in Figure~\ref{fig:1}(a)). In particular, the 60--150~keV flux decreased rapidly to the pre-flare level by $\sim$00:57 UT. Note that the HXI data were affected by radiation belt contamination in the interval of 00:58--01:04 UT, as indicated by the light gray shaded area. The profile of SXR time derivative showed a strong peak at 00:56~UT, consistent with those of the HXR emissions from HXI at the energy bands greater than 20 keV, corresponding to the Neupert effect \citep{1968ApJ...153L..59N}. Figure~\ref{fig:1}(b) shows the time profiles of WST 3600~\AA, HMI 6173~\AA, and AIA UV (1600 and 1700 \AA) emissions summed over the flare region (marked by the white dotted box in Figures~\ref{fig:1}(c) and (f)), with the pre-flare intensity subtracted. It is seen that all these emissions reached their peaks around 00:56~UT, demonstrating a temporal correspondence with the HXR peaks.

Figures~\ref{fig:1}(c)--(h) exhibit the base-difference images of WST and HMI together with the AIA 131, 171, and 304 \AA\ images at the flare peak time and during the gradual phase of the flare. Four footpoint-like WL sources, including two inner ones labeled as FP1 and FP2 and two outer ones labels as FP3 and FP4, were identified, all of which were still located on the solar disk. Note that some WL brightenings can also be seen above the solar limb, where a logarithmic transformation is applied in WST and HMI images to enhance the visibility. Here we use a box with a size of $6^{\prime\prime}\times6^{\prime\prime}$ to present each of the sources. The two inner sources of FP1 and FP2 exhibited substantially stronger enhancements than the outer ones of FP3 and FP4 (Figures \ref{fig:1}(c) and (d)). In addition, all the four footpoint sources were clearly visible in WST, while the outer sources appeared much fainter in HMI. These four WL sources root in a quadrupolar magnetic structure overall (Figure \ref{fig:1}(h)), though they have mixed magnetic polarities as seen from the HMI magnetogram (Figure \ref{fig:1}(i)), which is affected by projection effect due to close to the limb. This WLF was accompanied by a failed filament eruption, during which most of the filament material fell back toward the vicinity of the WL source of FP4, as revealed by the multiwavelength images in Figure~\ref{fig:1}(h) and the online animation.

\subsection{Morphology Evolution and Spatial Correlation between the WL and HXR Sources}
\label{subsec:hxr}

\begin{figure*}[ht!]
\plotone{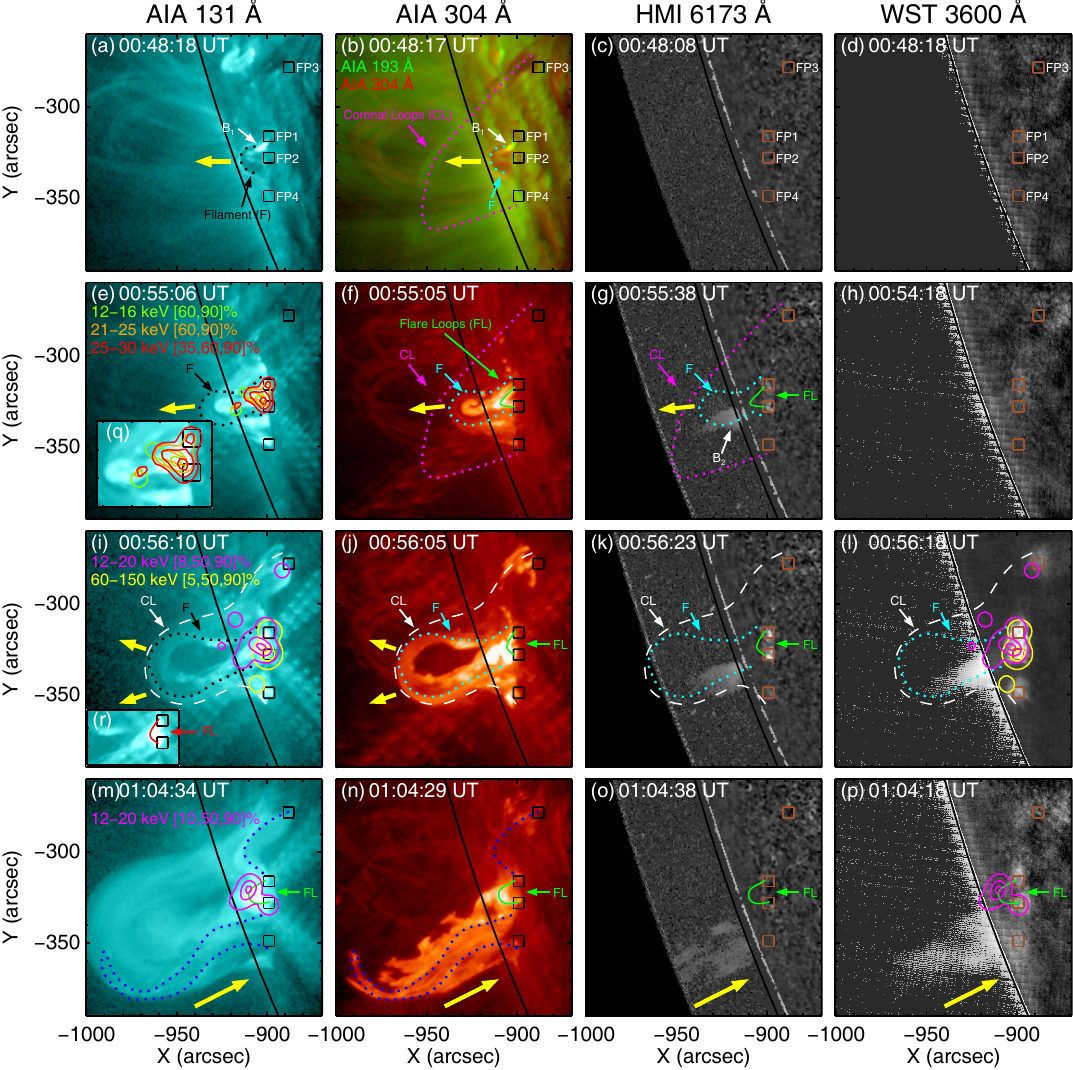}
\caption{Temporal evolution of the AIA 131~\AA\ images (first column), AIA 304~\AA\ images (second column), and base-difference images of HMI 6173~\AA\ (third column) and WST 3600~\AA\ (fourth column). The contours of HXI HXR sources are overplotted on the AIA 131~\AA\ and WST 3600~\AA\ images. The AIA 193~\AA\ emission is also shown in green in panel (b). The black or brown boxes in each panel denote the locations of four footpoint sources (FP1--FP4) as detected in WST and HMI images. The labels ``F", ``CL", ``FL" denote the filament, coronal loops, and flare loops in selected panels, respectively. The ``$\mathrm{B_1}$" in panels (a) and (b) indicates the brightenings beneath the filament and the ``$\mathrm{B_2}$" in panel (g) indicates the off-limb WL brightening. The thick yellow arrows indicate the eruption and subsequent motion of the filament material. The solar limb is outlined by black curves. (q) and (r) Zoomed-in views of the flare region in panels (e) and (i), respectively. An animation of the temporal evolution is available in the online Journal. In the animated view, the top-left panel shows the GOES SXR 1--8~\AA\ light curve, with three vertical black dotted lines representing the start, peak, and end times of the flare and the magenta dotted line indicating the time corresponding to the current animation frame. The remaining 11 panels display the AIA images in seven EUV and two UV passbands, as well as the base-difference images of WST 3600~\AA\ and HMI 6173~\AA. Four magenta boxes in each of these panels denote the locations of four footpoint sources (FP1--FP4) as detected in WST and HMI images.}
\label{fig:evo}
\end{figure*}

Figure~\ref{fig:evo} presents the temporal evolution of the AIA 131 \AA\ and 304 \AA\ images as well as the base-difference images of WST~3600 \AA\ and HMI~6173 \AA. The HXR sources observed by HXI are overplotted on AIA 131 \AA\ and WST 3600 \AA\ images to reveal the spatial relationship among the HXR, WL, and coronal sources during the flare. The AIA 304~\AA\ images could help trace the morphology and evolution of the erupting filament throughout the flare process, and the thick yellow arrows indicate the rising, expansion, and fallback motions of the filament material at different stages of the eruption.

At 00:48 UT (Figures \ref{fig:evo}(a)--(d)), in the early rise phase of the flare, some brightening (denoted as ``$\mathrm{B_1}$" in Figures \ref{fig:evo}(a) and (b)) appeared beneath the filament (marked by the black or cyan dotted line and denoted as ``F"), possibly associated with the slight enhancement of SXR 1--8 \AA\ flux. Subsequently, the filament gradually lifted and expanded (Figure \ref{fig:evo}(f) and the online animation), accompanied by a persistent brightening beneath it. The HXR emissions started to increase at 00:55 UT and two HXR footpoint sources showed up at FP1 and FP2, together with a cusp-like structure (see the HXR contours in Figures \ref{fig:evo}(e) and (q)). The two footpoint sources connected some reconnecting flare loops (marked by the green curve and denoted as ``FL" in Figure \ref{fig:evo}(f)) below the filament. At this time, two WL brightenings were also observed by HMI at FP1 and FP2 (Figure \ref{fig:evo}(g)). Besides, an HXR coronal source above the loop top appeared Figure \ref{fig:evo}(e)). Note that the coronal source at a relatively higher energy band (25--30 keV) is located slightly lower in altitude than that at a lower-energy band (12--16 keV or 21--25 keV) (Figure \ref{fig:evo}(q)), suggesting a magnetic reconnection beneath the rising filament \citep{2008ApJ...676..704L}. Some weak off-limb WL brightening (denoted as ``$\mathrm{B_2}$") also appeared at $\sim$00:55 UT in the HMI image (Figure \ref{fig:evo}(g)), likely corresponding to some filament material.

At 00:56~UT (Figures \ref{fig:evo}(i)--(l)), the filament underwent a rapid ascent and exhibited an $\Omega$ shape. It rose to the height of the overlying coronal loops, i.e., the large-scale coronal loops (denoted as ``CL") as seen in AIA 193 \AA\ images during the early rise phase of the flare (see the magenta dotted line in Figure \ref{fig:evo}(c) and also the online animation), and interacted with those loops. At this stage, two remote brightenings appeared at the two outer footpoints of FP3 and FP4, which were more clearer in the WST 3600 \AA\ wave band, as shown in Figures~\ref{fig:evo}(k) and (l). It should be mentioned that the inner WL sources of FP1 and FP2 further brightened as seen in the HMI images (Figures \ref{fig:evo}(g) and (k)). The existence of two coronal sources at 12--20 keV above the cusp-like structure (Figure \ref{fig:evo}(i)) might imply a magnetic reconnection between the erupting filament and the background coronal field. The post-reconnected loops can be seen in the AIA 131 \AA\ images or traced by the filament material in the AIA 304 \AA\ images a few minutes later (see the blue dotted lines in Figures \ref{fig:evo}(m) and (n)). Regarding all the four HXR footpoint sources, it is seen that the southern footpoints of FP2 and FP4 exhibited stronger HXR emissions than their northern counterparts of FP1 and FP3. In particular, FP4 corresponded to an HXR source at 60--150 keV, though having a little offset, while its counterpart of FP3 showed no detectable emission at such a high energy band.

Ultimately (after $\sim$00:58 UT), the filament experienced a failed eruption, possibly due to a strong confinement by the overlying magnetic field. It gradually fragmented, and a significant amount of material fell back to the region around FP4 (and FP2 as well, see the online animation), as indicated by the thick yellow arrow in Figure~\ref{fig:evo}(n), which can also be clearly visible in the HMI and WST images (Figures \ref{fig:evo}(o) and (p)). The HXI 12--20 keV source at the loop top between FP1 and FP2 remained present and extended towards FP2 at $\sim$01:04~UT (Figure \ref{fig:evo}(m)). Besides, higher-lying diffuse loops became visible, more prominently in the high-temperature 131 \AA\ wave band. 

\subsection{Evolution of the Inner Footpoints of FP1 and FP2}
\label{subsec:inneranalysis}

\begin{figure*}[ht!]
\plotone{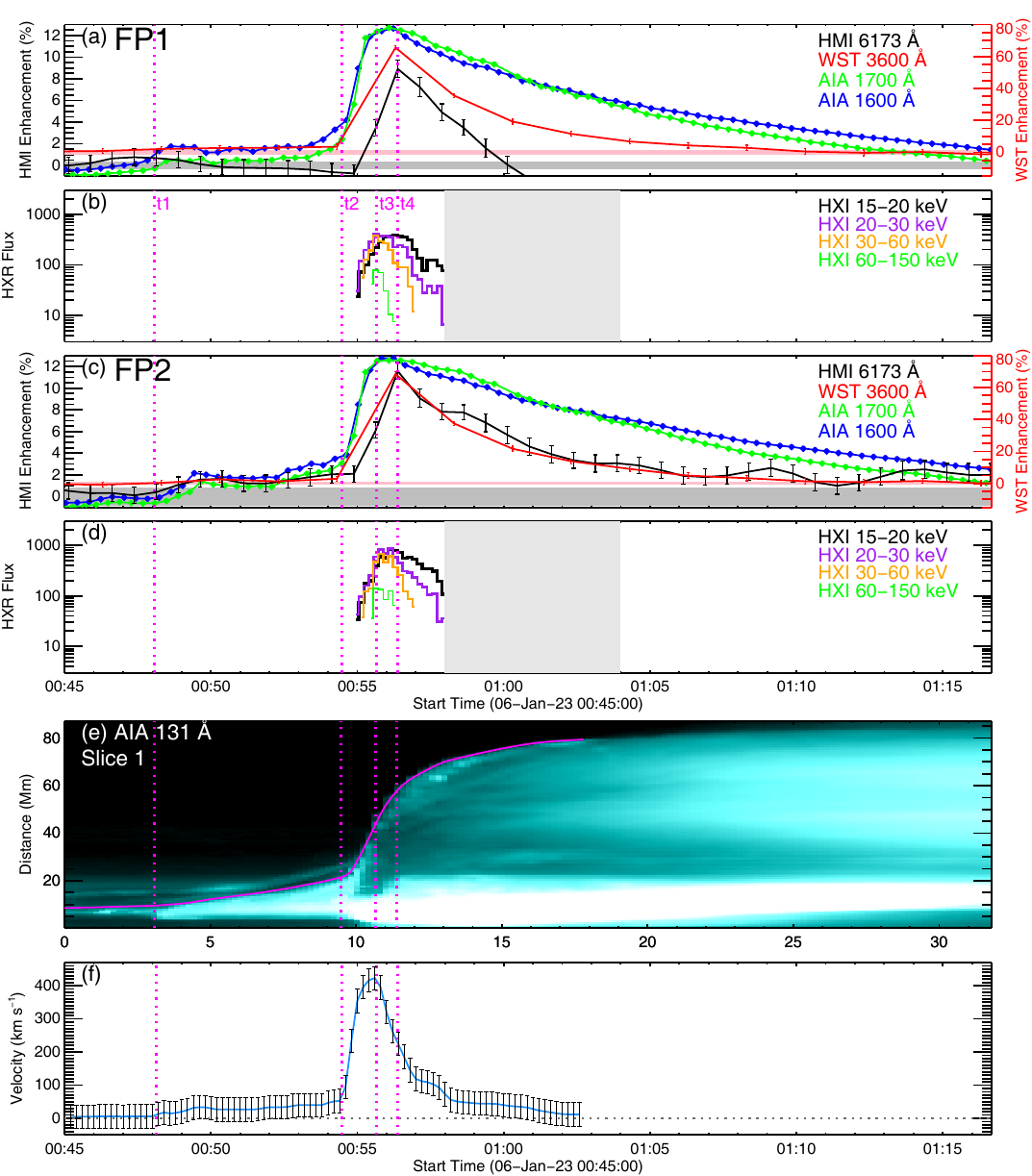}
\caption{(a) Enhancements of the HMI 6173~\AA\ (corresponding to the left coordinate) and WST 3600~\AA\ (corresponding to the right coordinate) emissions at FP1. The gray and pink horizontal shaded bars mark the uncertainties of HMI and WST emissions, respectively, estimated using the pre-flare data from 20 minutes before the flare event. The WST and HMI measurement error at each time point is alternatively estimated using data from the quiet-Sun region (indicated by the orange dotted box in Figure~\ref{fig:1}(f)). The four vertical magenta dotted lines (also in panels (b)--(f)) indicate the times of t1, t2, t3, and t4. (b) Light curves of HXI 15--20 keV, 20--30 keV, 30--60 keV, 60--150 keV emissions integrated over FP1, in units of $\mathrm{photons\ cm^{-2}\ s^{-1}\ arcsec^{-2}}$. The light gray shaded area marks the period when HXI data are affected by the radiation belt. (c) and (d) Similar to panels (a) and (b), but for FP2. (e) Time-distance plot along Slice~1 (indicated in Figure~\ref{fig:1}(e)) based on AIA 131 \AA\ images. The magenta solid curves trace the leading edge of the filament. (f) The blue curve shows the filament velocity, with the uncertainties marked by black short lines. The horizontal gray dotted line indicates the velocity of zero. 
\label{fig:lc12}}
\end{figure*}

To better understand the nature of the WL emissions, we first investigate the temporal evolution of the two inner footpoint sources of FP1 and FP2, as shown in Figure~\ref{fig:lc12}. Figures~\ref{fig:lc12}(a)--(d) display the multiwavelength emission curves at FP1 and FP2, i.e., integrated over the box of 6$^{\prime\prime} \times$6$^{\prime\prime}$ for each wave band. It should be mentioned that this box size is similar to the spatial resolution of the HXI imaging for this flare. The WST 3600 \AA\ and HMI 6173 \AA\ continuum enhancements are calculated by $(I-I_{0})/I_{0}$, where $I$ and $I_{0}$ represent the intensity at the flare time and the mean pre-flare intensity over 00:30--00:40 UT, respectively. The time-distance plot at AIA 131 \AA\ along Slice 1 (marked in Figure \ref{fig:1}(e)) between FP1 and FP2 is shown in Figure~\ref{fig:lc12}(e), and velocities of the related filament derived from the slice plot are exhibited in Figure~\ref{fig:lc12}(f). 

At $\sim$00:48~UT (t1, indicated by the leftmost vertical dotted line in Figure \ref{fig:lc12}), the filament’s ascent velocity increased slightly (Figure~\ref{fig:lc12}(f)) and a distinct brightening appeared beneath the filament (Figure~\ref{fig:lc12}(e)). Note that the leading edge of the filament is marked as a magenta curve in Figure~\ref{fig:lc12}(e). Simultaneously, the AIA UV 1600 and 1700~\AA\ emissions at FP1 and FP2 gradually intensified (Figures~\ref{fig:lc12}(a) and (c)). Around 00:54~UT (t2), the UV emissions at FP1 and FP2 rapidly increased. In the mean time, the WL emissions at both footpoints began to exhibit significant enhancements, corresponding to the rapid rise of the filament (Figures~\ref{fig:lc12}(e) and (f)). About 40~s later (shortly before 00:56~UT, t3), the higher-energy HXR emissions at 30--60~keV and 60--150~keV reached their peaks (Figures~\ref{fig:lc12}(b) and (d)). The filament’s eruption velocity also reached its maximum of $\sim$420~km~s$^{-1}$. This timing corresponded to the rise phase of the WST and HMI continuum emissions. At $\sim$00:56~UT (t4), the lower-energy HXR emission of 15--20 keV reached its peak. The WST and HMI continuum emissions at both footpoints reached their maxima as well. The maximum enhancements of WST and HMI continua are 65.0\% and 8.9\% at FP1, respectively, and slightly smaller than those (68.5\% and 11.6\%, respectively) at FP2. Note that the enhancement at FP1 drops below zero after $\sim$01:01 UT, mainly due to a decreasing intensity of the sunspot umbra, where FP1 is located, compared with the pre-flare level. It should be mentioned that the HXR emissions at FP1 and FP2 displayed a more asymmetric pattern in the peak flux. Specifically, the peak HXR fluxes at FP1 at 15--20~keV and 60--150~keV reached 79 and 391 photons~cm$^{-2}$~s$^{-1}$~arcsec$^{-2}$, respectively, while FP2 attained higher peak fluxes of 142 and 825 photons~cm$^{-2}$~s$^{-1}$~arcsec$^{-2}$ in the corresponding bands, i.e., about twice the values at FP1.  

\subsection{Evolution of the Outer Footpoints of FP3 and FP4}
\label{subsec:outeranalysis}

\begin{figure*}[ht!]
\plotone{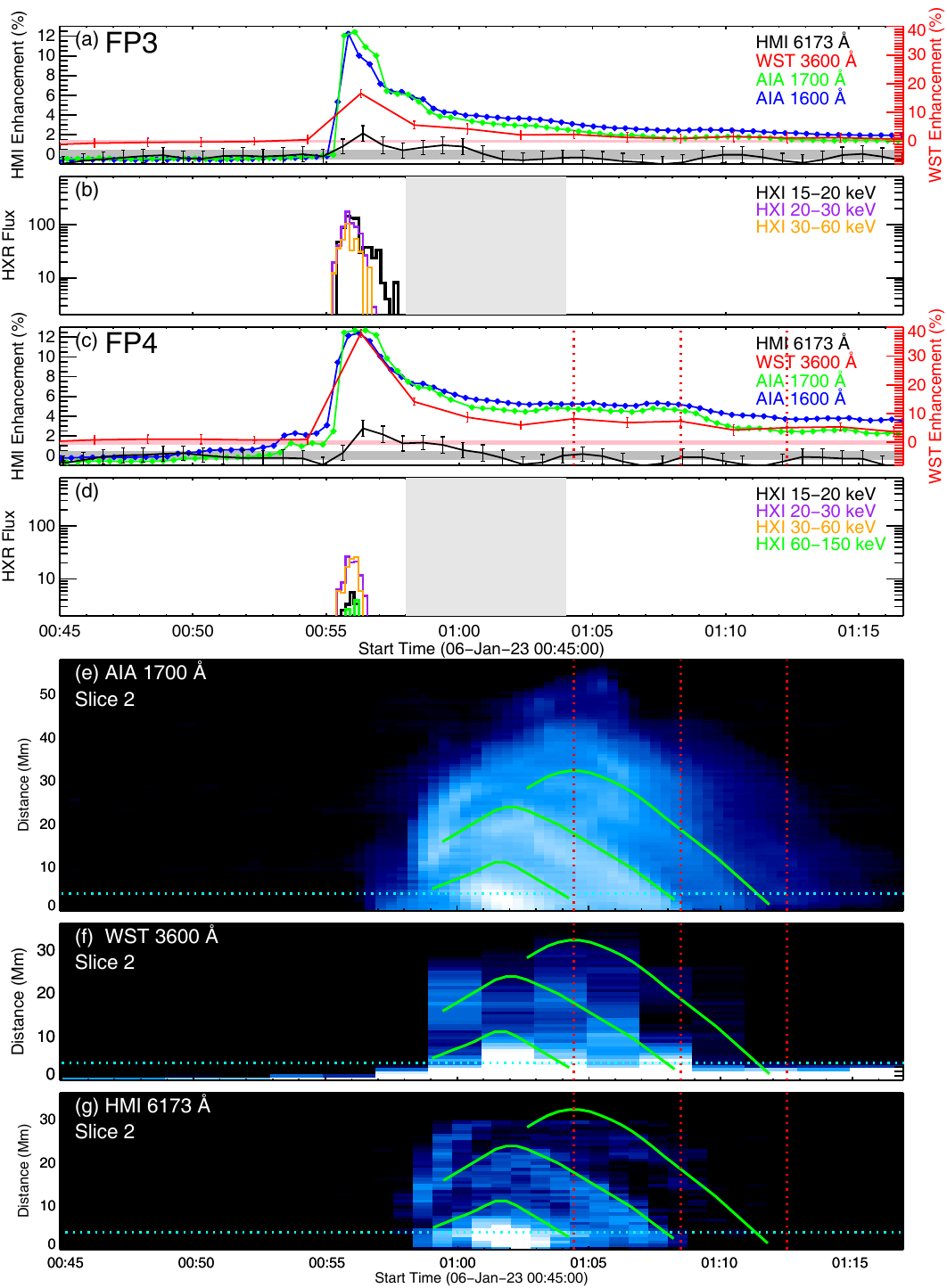}
\caption{(a)--(d) Similar to Figures \ref{fig:lc12}(a)--(d), but for FP3 and FP4. (e)--(g) Time-distance plots of AIA 1700 \AA, WST 3600 \AA, and HMI 6173 \AA\ emissions along Slice 2, as indicated in Figure~\ref{fig:1}(h), which are processed by taking the logarithm and applying linear enhancement in the Y-direction to highlight the bright structures. The three green curves in each panel trace the evolution of three bright structures, and the three red vertical dotted lines (also in panel (c)) mark their respective end times. The cyan horizontal dotted lines indicate the position of the solar limb.
\label{fig:lc34}}
\end{figure*}

Figures~\ref{fig:lc34}(a)--(d) show the temporal profiles of multiwavelength emissions at the outer WL sources of FP3 and FP4. In addition to the prominent peak around 00:56 UT, it is notable that the WST 3600~\AA\ emission together with the AIA UV emissions (especially at FP4) still persist in the late phase of the flare (say after 01:04 UT), which are actually associated with the fallback of filament material (Figures~\ref{fig:lc34}(e)--(g)).

Similar to the inner sources of FP1 and FP2, the outer FP3 and FP4 sources exhibited brightenings in the WL (3600 and 6173 \AA) and UV (1600 and 1700 \AA) wave bands, accompanied by HXR emissions. In particular, both WST 3600 \AA\ and HMI 6173 \AA\ emissions at FP3 and FP4 peaked around 00:56~UT, at the same time as those of the two inner footpoints. However, the multiwavelength emissions showed some differences between FP3 and FP4. Firstly, the maximum enhancements of WST 3600~\AA\ emission are 16.6\% at FP3 and 38.0\% at FP4, displaying a strong asymmetry between the two footpoints (Figures \ref{fig:lc34}(a) and (c)). By contrast, the maximum enhancements of HMI 6173 \AA\ emission are 2.1\% and 2.8\% at FP3 and FP4, respectively, i.e., similar with each other, though much lower than those of 3600 \AA\ emissions. Secondly, the UV 1600 and 1700~\AA\ brightenings began around 00:55~UT at FP3, while they started near 00:53~UT, i.e., two minutes earlier, at FP4. Note that both the UV brightenings at FP3 and FP4 appeared later than those at FP1 and FP2, which commenced at approximately 00:48~UT. Finally, the HXR emissions showed different features at FP3 and FP4. FP4 exhibited a distinct response in the 60--150~keV band (Figure \ref{fig:lc34}(d), also see Figure~\ref{fig:evo}(i)), whereas no such signal was detected at FP3 (Figure \ref{fig:lc34}(b)). The bombardment by more energetic nonthermal electrons at FP4 may account for its stronger enhancement in the WST 3600 \AA\ continuum. 

Interestingly, a plateau in the AIA UV light curves and some weak bumps in the WST continuum emission were observed at FP4 during the late decay phase between $\sim$01:04 and $\sim$01:16~UT, as indicated by the vertical red dotted lines in Figure~\ref{fig:lc34}(c). The HMI continuum also exhibits marginal enhancements during this period, although their enhancements remain within the uncertainty level. Figures~\ref{fig:lc34}(e)--(g) present the time-distance plots of AIA 1700~\AA, WST~3600~\AA, and HMI 6173~\AA\ emissions along Slice 2 that crosses FP4 (indicated in Figure~\ref{fig:1}(h)). The three green curves trace the trajectories of some ejected and returning materials, and the timings of material return to the solar surface correspond well to the weak enhancements of AIA UV and WST emissions and maybe HMI emission as well. This suggests that, as the filament material fell back, its kinetic energy converted into thermal energy upon impact, thereby heating the lower atmosphere at FP4 and producing the observed UV and WL enhancements. Note that the quasi-periodic fluctuations with a period of $\sim$5~minutes in the HMI 6173 \AA\ emission at FP4 may also be caused by some intrinsic oscillations, such as p-mode oscillations \citep[e.g.,][]{1962ApJ...135..474L, 2002RvMP...74.1073C}, since in addition to FP4, the other three footpoint sources display some similar fluctuations in the HMI intensity (Figures \ref{fig:lc12}(c) and \ref{fig:lc34}(a)).

\subsection{Comparison of the WL and HXR Paremeters among FP1--FP4}

To investigate the relationship between the HXR emission and WL enhancement at different footpoints, we made an analysis on the spatially resolved HXR fluxes at the four footpoints around the flare peak time ($\sim$00:55:48 UT). Specifically, we derived the HXR fluxes ($F$) in six energy bands of 26--30, 30--34, 34--38, 38--42, 42--46, 46--50~keV ($E$) above the estimated low-energy cutoff of $\sim$20~keV (see the spectral fitting results in \citealt[][]{2024SoPh..299...30S} for the same flare event, their Figure 5) for each footpoint source and fitted them with a power-law function ($F \propto E^{-\gamma}$) to obtain the photon spectral index ($\gamma$) \citep{2023A&A...670A..89S}. The fitting result is shown in Figure \ref{fig:hxi_at_fourfps}. Under the thick-target approximation \citep[e.g.,][]{1975SoPh...41..135B}, the corresponding electron spectral index can be estimated as $\delta \approx \gamma + 1$, which are listed in Table \ref{tab:fp}. We can see clear differences among the four footpoints: FP1 and FP2 exhibit relatively hard-to-intermediate spectra with $\delta \approx 3.7$ and $4.2$, respectively, while FP3 shows a softer spectrum ($\delta \approx 5.1$). FP4, despite having the lowest HXR flux, presents the hardest spectrum among the four footpoints ($\delta \approx 3.6$). In terms of the peak WL enhancement (also listed in Table \ref{tab:fp}), FP1 and FP2 show the strongest enhancements, while FP3 and FP4 are significantly weaker.

\begin{deluxetable*}{lccccc}[h!]
\tablecaption{ HXR and WL Parameters of the Four Footpoints\label{tab:fp}}
\tablehead{
\colhead{} &
\colhead{HXI Peak Counts} &
\colhead{$\gamma$} &
\colhead{$\delta\approx\gamma+1$} &
\colhead{WST enhancement} &
\colhead{HMI enhancement}
}
\startdata
FP1 & 1253.1 & 2.7 & 3.7 & 65.0\% & 8.9\% \\
FP2 & 2486.0 & 3.2 & 4.2 & 68.5\% & 11.6\% \\
FP3 & 426.0  & 4.1 & 5.1 & 16.6\% & 2.1\% \\
FP4 & 61.5   & 2.6 & 3.6 & 38.0\% & 2.8\% \\
\enddata
\tablecomments{The HXI peak counts are calculated as the sum of the peak counts in the four HXI energy bands (15--20, 20--30, 30--60, and 60--150 keV), in units of $\mathrm{photons\ cm^{-2}\ s^{-1}\ arcsec^{-2}}$.}
\end{deluxetable*}

According to radiative hydrodynamic simulations \citep[e.g.,][]{2015ApJ...809..104A, 2026A&A...705A.157O}, the WL response are controlled not only by the energy flux but also by the spectral index of nonthermal electron beams when the low-energy cutoff is fixed. A harder spectrum (having a lower spectral index) allows a larger fraction of high-energy electrons to penetrate deeper into the lower atmosphere, increasing the efficiency of WL production, and a higher energy flux leads to a stronger overall heating. For the present WLF, the strong WL enhancements at FP1 and FP2 are primarily driven by their substantially higher HXR fluxes, whereas the relatively stronger WL emission at FP4 compared to FP3 can be attributed to its harder HXR spectrum. These results confirm that both factors of energy flux and spectral index are important, with their relative contributions varying among different footpoints. In particular, this naturally explains the discrepancy between FP3 and FP4, where the harder spectrum at FP4 compensates for its lower HXR flux and leads to a relatively stronger WL enhancement.

\section{Summary and Discussions} \label{sec:summary}

In this paper, we present a rare quadrupolar WL source configuration observed by ASO-S/LST/WST in an X1.2 flare that was associated with a failed filament eruption. Combining with some multiwavelength observations from SDO/HMI, SDO/AIA, and ASO-S/HXI, it is revealed that the two inner WL sources share similar characteristics, while the two outer ones show some differences. Our main results are summarized as follows.

\begin{enumerate}

\item Four flare-related footpoint sources (FP1--FP4) were identified in both the WST 3600~\AA\ and HMI 6173~\AA\ continuum images. These quadrupolar WL sources were still located on the solar disk, though close to the limb. Some off-limb WL brightenings were also observed, mainly caused by the filament eruption and plasma fallback. 

\item The four WL footpoint sources appeared nearly simultaneously as seen from the WST and HMI images, whereas in the UV and EUV wave bands, the two inner footpoints of FP1 and FP2 brightened earlier than the two outer ones of FP3 and FP4. In addition, the relative enhancements of 3600/6173~\AA\ emissions at the inner footpoints (65.0\%/8.9\% for FP1 and 68.5\%/11.6\% for FP2) are significantly higher than those at the outer footpoints (16.6\%/2.1\% for FP3 and 38.0\%/2.8\% for FP4). 

\item The four WL sources spatially corresponded to the HXR footpoint sources at different energy bands. In particular, the two inner WL/HXR footpoint sources of FP1 and FP2 connected a cusp-like structure with a loop-top source above it as seen in the HXR imaging, suggesting a magnetic reconnection occurring during the flare event. 

\item In addition to the relatively strong WL brightenings during the impulsive phase of the flare at FP1--FP4, some weak WL brightenings showed up at FP4 during the late decay phase, which were probably caused by the fallback of filament material, though some intrinsic oscillations cannot be excluded for the HMI 6173 \AA\ emission.
\end{enumerate}

From the above analysis, we can infer the relationships and differences among the four WL footpoint sources. The two inner sources (FP1 and FP2) are a pair of footpoints of some flare loops, while the two outer sources (FP3 and FP4) exhibit different observational characteristics, corresponding to the erupting filament. Despite the nearly simultaneous WL brightenings of the four footpoint sources, likely due to the limited temporal resolution, the AIA EUV and UV imaging observations suggest a sequential process: an initial slow reconnection in the core field leads to an accelerated rise of the filament, which is also reflected in the early off-limb brightenings as seen in HMI images. This rising motion subsequently triggers a magnetic reconnection between the filament and the overlying coronal loops or magnetic fields. The internal eruption appears to commence prior to the brightenings of the outer footpoint sources, likely supporting the tether-cutting model \citep{2006GMS...165...43M}. However, it should be noted that we cannot rule out the possibility that the eruption was triggered by ideal MHD instabilities or a loss of equilibrium.

Magnetic reconnection produces a substantial population of high-energy nonthermal electrons, with the highest responding energy band of above 60~keV for the WLF studied here, which show good temporal and spatial correspondence with the observed WL enhancements. The HMI 6173~\AA\ emission primarily originates in the photosphere, while the WST 3600~\AA\ continuum is mainly formed in the lower to middle chromosphere \citep[e.g.,][]{1986lasf.conf..216A, 1995A&AS..110...99F}, where electrons with energies of 50--100~keV can effectively deposit the energy \citep{2015ApJ...809..104A}. So the electron precipitation can directly produce the 3600 \AA\ emissions, while the secondary processes such as radiative backwarming may also contribute indirectly. The stronger WL enhancements at the inner footpoints are likely due to the presence of more numerous and higher-energy nonthermal electrons compared to the outer footpoints. During the filament eruption, the cool material could also contribute to the off-limb coronal WL brightening. In addition, due to the failed eruption, a large amount of plasma falls back and causes the gradual-phase brightening at FP4, especially in the 3600 \AA\ continuum. 

Due to different formation heights in the solar atmosphere, the WST 3600~\AA\ emission in the Balmer continuum and the HMI 6173~\AA\ emission in the Paschen continuum exhibit some different features in this X1.2 WLF. In the traditional classification, WLFs are divided into Type I and Type II based on observational features such as the relative enhancement between the Balmer and Paschen continua (i.e., the presence or absence of the Balmer jump). In this large WLF, the WL and HXR emissions exhibit a good temporal and spatial correlation at the footpoints, and the enhancement at 3600~\AA\ is significantly greater than that at 6173~\AA. These indicate that this is a Type I WLF. Considering that the 3600 \AA\ emission is more sensitive to the nonthermal electron beam heating compared with the 6173 \AA\ emission \citep{2024ApJ...972L...1L}, it is necessary to combine these two continua to comprehensively understand the physical mechanisms of solar WLFs. 

\begin{acknowledgments}
We thank the referee for his/her careful evaluation of our manuscript, whose constructive comments and suggestions greatly improved the quality and clarity of the manuscript. The ASO-S mission is supported by the Strategic Priority Research Program on Space Science, the Chinese Academy of Sciences. SDO is a mission of NASA’s Living With a Star Program. We thank Zhentong Li, Wenhui Yu, and Lingfang Wang very much for their helpful discussions. This work was supported by the Strategic Priority Research Program of the Chinese Academy of Sciences under grant XDB0560000, the National Key R\&D Program of China under grant 2022YFF0503004, and NSFC under grants 12273115 and 12233012. D.-C.S. is supported by the Jiangsu Funding Program for Excellent Postdoctoral Talent and the China Postdoctoral Science Foundation under Grant Number 2025M773193.
\end{acknowledgments}
\vspace{5mm}

\appendix
\section{HMI six-wavelength-point spectra at the four footpoints}\label{sec:appendix}
We use the HMI 720 s cadence spectral data to examine the enhancements of the \ion{Fe}{1} 6173~\AA\ line and the nearby continuum at the four footpoints. The available times include $t_a=$ 00:47 UT, corresponding to the early rise phase of the flare, and $t_b=$ 00:59 UT, two minutes after the flare peak. Figure \ref{fig:hmi_spectrum} shows the observed six-wavelength-point spectra of \ion{Fe}{1} (black and red asterisks) at $t_a$ and $t_b$, together with the corresponding Gaussian fits (black dotted and red solid curves). It can be found that the line profiles at all four footpoints remain in absorption during the flare and exhibit an overall intensity enhancement from $t_a$ to $t_b$, with FP1 and FP2 showing stronger increases compared to FP3 and FP4. In particular, some noticeable enhancements are present in the continuum close to the red wing for FP3 and FP4. These results indicate that the WL enhancements originate from the combined contribution of all spectral sampling points, including the nearby continuum. This also supports the reliability of the reconstructed HMI continuum for the present event.

\bibliography{refer}{}
\bibliographystyle{aasjournal}

\begin{figure*}[ht!]
\plotone{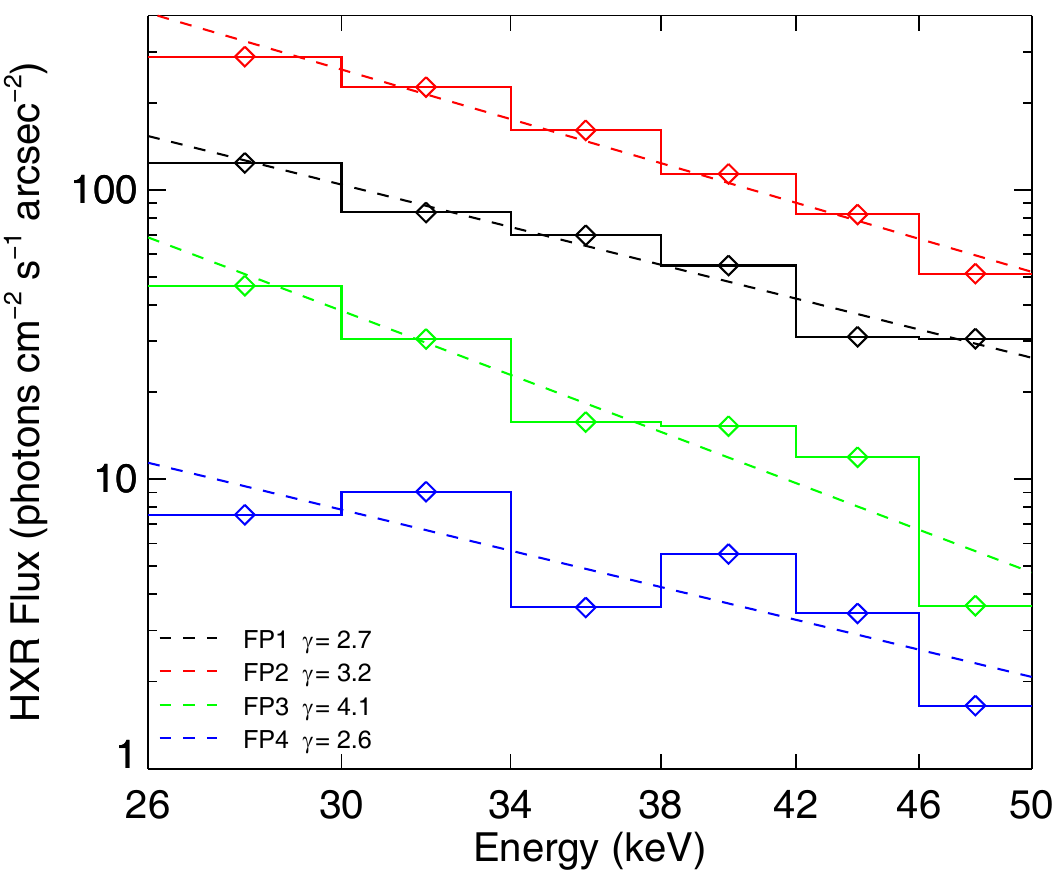}
\caption{Power-law fits to the spatially resolved HXR fluxes measured in six energy bands between 26 and 50~keV for the four WL sources of FP1--FP4. The diamonds represent the observed HXR fluxes in each energy band and the dashed lines are the fitting results. Different colors denote different WL sources, with the corresponding power-law indices ($\gamma$) indicated.}
\label{fig:hxi_at_fourfps}
\end{figure*}

\begin{figure*}[ht!]
\renewcommand{\thefigure}{A\arabic{figure}} 
\setcounter{figure}{0}    
\renewcommand{\theHfigure}{A\arabic{figure}}
\plotone{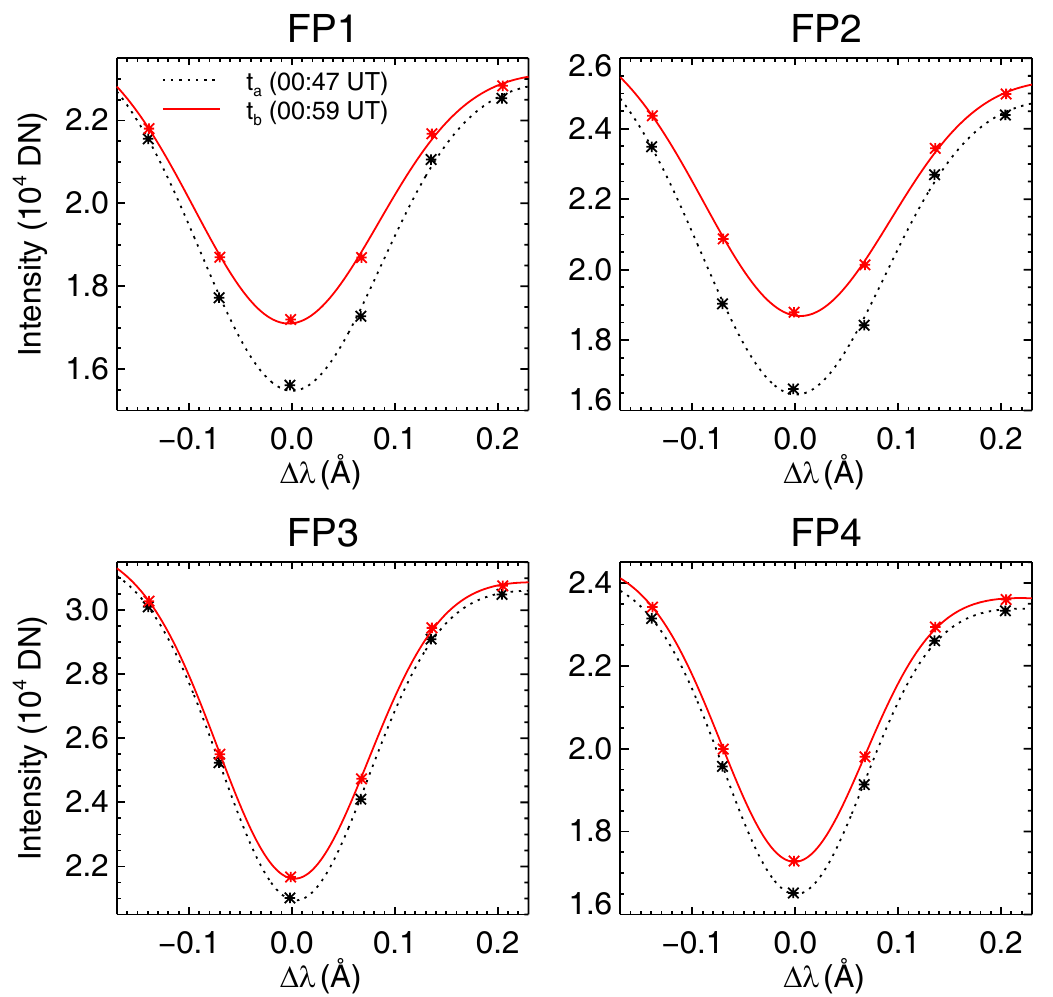}
\caption{HMI six-wavelength-point spectra of \ion{Fe}{1} 6173 \AA\ at FP1--FP4 in the early rise phase of the flare ($t_{a}=$ 00:47 UT, black asterisks) and two minutes after the flare peak ($t_{b}=$ 00:59 UT, red asterisks), together with the Gaussian fits (black dotted and red solid curves). The horizontal axis denotes the wavelength relative to the line center, which is calibrated via a quiet-Sun region close to the flare region.}
\label{fig:hmi_spectrum}
\end{figure*}

\end{document}